\documentclass[12pt]{article}
\title{Coherent states in complex variables $SU(2S+1)/SU(2S)\bigotimes U(1)$   and classical dynamics }
\author{Y.Yousefi; Kh. Kh. Muminov \\
Physical-Technical Institute named after S.U.Umaro\\
 Academy of Sciences of Republic of Tajikistan\\
Aini Ave 299/1, Dushanbe, Tajikistan\\
e-mail: yousof54@yahoo.com; Khikmat@inbox.ru}
\date{}
\begin{document}
\maketitle
\begin{abstract}
It was studied coherent states in complex variables in SU(2), SU(3), SU(4) groups and in general in SU(n) group.  Using the completeness relation of the coherent state, we obtain a path integral expression for transition amplitude which connects a pair of SU(n) coherent states. In the classical limit, a canonical equation of motion is obtained.
\end{abstract}
\section{Introduction}
One of the main motives for the use of Feynman’s path integral in quantum mechanics lies in its initiative way of describing the correspondence between classical and quantum concepts. Especially the integration over paths in phase space gives Hamiltonian’s equation of motion in the classical limit. According to this, the system is firstly supposed to propagate through infinite sequence of coordinate eigenstate, and then via the transformation to momentum representation at each time interval the transition amplitude is brought into the form of integration over the paths in phase space. There is, however, another way of deriving the phase space path integral through the introduction of the coherent state. 

In the quantum mechanics a coherent state (hereafter abbreviated as CS) is a specific kind of quantum state of the quantum harmonic oscillator whose dynamics most closely resemble the oscillating behavior for a classical harmonic oscillator system. The most important properties of coherent state are the continuity and completeness. As the ordinary CS is closely related with the unitary representation of Heisenberg-Weyl group, so the generalized coherent state has been introduced by Perelomov [1] in related to the unitary representation of an arbitrary Lie group. Section 2 is devoted to the properties, Casimir operator, path integral expression for the transition amplitude and the classical equation of the motion in  SU(2) group. There is similar expression  in section 3 for SU(3) group, in section 4 for SU(4) group and in section 5 for SU(2S+1) group.

In condensed matter physics, coherent states for  SU(n) group have extensively used to study Heisenberg or Non-Heisenberg spin systems using the path integral formalism.

\section{Properties of the SU(2) coherent state and classical dynamics}

According to the Ref (1), the generalized CS is given by the set  $ U(g)|0\rangle , g\in G$ , where U(g) is the unitary representation  of the lie group G acting on a Hilbert space and $|0\rangle$   is a fixed vector in this space.

In the case G=SU(2), U(g) can be parameterized as the following form [1,2]:

\begin{eqnarray}
|\psi\rangle=e^{(\alpha S^+-\bar \alpha S^-)}|0\rangle=(1+|\xi|^2)^{-J}e^{\xi S^+}|j,-j\rangle
\end{eqnarray}

Where the complex variable $\alpha$   takes the value in $|\alpha|\le \frac{\pi}{2}$   and the new parameter $ \xi$   takes an arbitrary value in the complex plane, and is related to $\alpha$  through:

\begin{eqnarray}
\xi=\frac{\alpha}{|\alpha|}tan|\alpha|
\end{eqnarray}

  $ S_i $ are generators of SU(2) group, and  are related to the Pauli matrices. The Pauli matrices are:

\begin{eqnarray}
\sigma_1=
 \left(
\begin{array}{cc}
0 & 1 \\
1 & 0  
\end{array}
\right), \;
\sigma_2=
 \left(
\begin{array}{cc}
0 & -i \\
i & 0  
\end{array}
\right), \;
\sigma_3=
 \left(
\begin{array}{cc}
1 & 0 \\
0 &-1  
\end{array}
\right) \;
\end{eqnarray}

The quadratic operator (Casimir operator) is defined in the following form

\begin{eqnarray}
\hat C_2=(S^x)^2+(S^y)^2+(S^y)^2=(S^z)^2+\frac{1}{2}(S^+S^-+S^-S^+)
\end{eqnarray}

And averaged value of this operator is: 

\begin{eqnarray}
\hat C_2=s(s+1)\hat I & & {  } for s=\frac{1}{2}
\end{eqnarray}

Lets consider a Hamiltonian $\hat H$   acting in our Hilbert space. Assume that $\hat H$ can be expanded as the finite polynomial of the infinitesimal operators $ \hat S^{\pm}, \hat S_z$ of SU(2). The transition amplitude (propagator) from state $ |\xi \rangle$  at time t to the state $|\xi^{'}\rangle$   at time $ t^{'}$   is given by 

\begin{eqnarray}
T(\xi^{'},t^{'},\xi,t)=\langle\xi^{'}|exp(-\frac{i}{\bar h}\hat H (t^{'}-t))|\xi \rangle
\end{eqnarray}

In order to obtain the path integral form  amplitude T, $(t^{'}-t)$  is divided into n equal time intervals $ \epsilon=\frac{(t^{'}-t)}{n}$  and take the limit $ n\rightarrow \infty$ :

\begin{eqnarray}
T=lim_{n\rightarrow\infty }\langle\xi^{'}|(1-\frac{i}{\bar h}\hat H\epsilon)^n | \xi \rangle
\end{eqnarray}

Using completeness relation and some mathematical calculation, we can get the following relation:

\begin{eqnarray}
T(\xi^{'},t^{'};\xi,t)&=&lim_{n\rightarrow\infty}\int...\int\prod_{k=1}^{n-1} d\mu(\xi_k) \nonumber\\
& &\times exp(\frac{i}{\bar h}\sum_{k=1}^n \epsilon (\frac{iJ\bar h}{1+|\xi_k|^2}(\xi_k^* \frac{\triangle \xi_k}{\epsilon}-\xi_k \frac{\triangle \xi_k^*}{\epsilon})-\langle\xi_k|\hat H|\xi_k \rangle)) \nonumber\\
& &
\end{eqnarray}

Below is presented the “formal” functional integral of the above expression:

\begin{eqnarray}
T&=&\int d\mu(\xi)exp(\frac{i}{\bar h}S) \nonumber\\
S&=& \int_t^{t{'}}L(\xi(t), \xi_t (t), \xi^*(t), \xi_t^*(t))dt
\end{eqnarray}

Where lagrangian L is given by 

\begin{eqnarray}
L=i(\frac{J\bar h}{(1+|\xi|^2)}(\xi^* \xi_t-\xi_t^* \xi)-\langle\xi|\hat H|\xi \rangle
\end{eqnarray}

In order to obtain classical equation of motion, the following condition should be used:

\begin{eqnarray}
0=\delta S&=& \int_t^{t{'}}(\frac{\partial L}{\partial\xi}\triangle \xi+\frac{\partial L}{\partial\xi_t}\triangle \xi_t+c.c.) \nonumber\\
&=& \int_t^{t{'}}((\frac{\partial L}{\partial\xi}-\frac{d}{dt}(\frac{\partial L}{\partial\xi_t^*}))\delta \xi+c.c.) dt
\end{eqnarray}

Because variations $ \xi$  and $\xi^*$  are independent and arbitrary, then
\begin{eqnarray}
\frac{d}{dt}(\frac{\partial L}{\partial\xi_t})-\frac{\partial L}{\partial\xi}=0, & &{  }\frac{d}{dt}(\frac{\partial L}{\partial\xi_t^*})-\frac{\partial L}{\partial\xi^*}=0
\end{eqnarray}

If L relation is used in the above equations, the classical equations take the following forms:

\begin{eqnarray}
\xi_t&=&-i\frac{(1+\xi^2)^2}{2J\bar h} \frac{\partial \langle\xi|H|\xi\rangle}{\partial \xi^*} \nonumber\\
\xi_t^*&=&i\frac{(1+\xi^2)^2}{2J\bar h} \frac{\partial \langle\xi|H|\xi\rangle}{\partial \xi}
\end{eqnarray}

\section{Properties of the SU(3) coherent state and classical dynamics}

Similar to equation (1), the SU(3) CS is written in the following form[3]

\begin{eqnarray}
|\psi\rangle=exp(\sum_{i=1}^{2}(\xi_i T_i^+-\bar \xi_i T_i^-))|0\rangle=(1+\sum_i^2|\psi_i|^2)^{-1/2}(|0\rangle+\sum_i^2\psi_i|i\rangle)
\end{eqnarray}

Where $ T_i$  are generators of SU(3) group and  related to Gell-Mann matrices. These matrices are

\begin{eqnarray}
\Lambda_1 &=&
 \left(
\begin{array}{ccc}
0 & 0&0 \\
0 & 0&-i \\
0&i&0  
\end{array}
\right), \;
\Lambda_2=
 \left(
\begin{array}{ccc}
0 & 0 &i \\
0& 0 &0 \\
-i& 0 &0 
\end{array}
\right),\; \nonumber\\
\Lambda_3&=&
 \left(
\begin{array}{ccc}
0 & -i&0 \\
i & 0&0 \\
0&0&0  
\end{array}
\right), \;
\Lambda_4=
 \left(
\begin{array}{ccc}
0 & 0&0 \\
0& 0&1 \\
0&1&0  
\end{array}
\right), \; \nonumber\\
\Lambda_5&=&
 \left(
\begin{array}{ccc}
0 & 0&1 \\
0 & 0&0 \\
1&0&0  
\end{array}
\right), \;
\Lambda_6=
 \left(
\begin{array}{ccc}
0 & 1&0 \\
1& 0&0 \\
0&0&0  
\end{array}
\right), \; \nonumber\\
\Lambda_7&=&\frac{1}{\sqrt{3}}
 \left(
\begin{array}{ccc}
1& 0 &0\\
0& 1 &0 \\
0& 0 &2 
\end{array}
\right),\; 
\Lambda_8=
 \left(
\begin{array}{ccc}
1 &0&0 \\
0& -1&0 \\
0&0&0  
\end{array}
\right), \;
\end{eqnarray}

 Coherent state is 

\begin{eqnarray}
|\psi\rangle=(1+\psi_1^2+\psi_2^2)^{1/2}(|0\rangle+\psi_1|1\rangle+\psi_2|2\rangle)
\end{eqnarray}

These states are parameterized by two complex functions  $\psi_1$ and $\psi_2$  , so the system lived on a four-dimensional real manifold. Where 

\begin{eqnarray}
\psi_i=\frac{\xi}{|\xi|}tan|\xi| &  & {  } |\xi|=\sqrt{\sum_{i=1}^2 |\xi_i|^2}, i=1,2
\end{eqnarray}

Similar to SU(2) group, the quadratic operator (Casimir operator) is  in the following form

\begin{eqnarray}
\hat C_2=(S^z)^2+\frac{1}{2}(S^+S^-+S^-S^+)=Q^{zz}+\frac{1}{2}(Q^{+-}+Q^{-+})
\end{eqnarray}

That $ Q^{zz}=\langle \psi| \hat S^z \hat S^z| \psi \rangle$  , $ Q^{+-}=\langle \psi| \hat S^+ \hat S^-| \psi \rangle$.The averaged Casimir  operator is 

\begin{eqnarray}
\hat C_2=s(s+1)\hat I, s=1
\end{eqnarray}

 The transition amplitude (propagator) from state $ |\psi \rangle$  at time t to the state $|\psi^{'}\rangle$   at time $ t^{'}$   is given by 

\begin{eqnarray}
T(\psi^{'},t^{'},\psi,t)&=&\langle\psi^{'}|exp(-\frac{i}{\bar h}\hat H (t^{'}-t))|\psi \rangle \nonumber\\
&=&lim_{n\rightarrow\infty}\int...\int\prod_{k=1}^{n-1} d\mu(\psi_k) exp(\frac{i}{\bar h}\sum_{k=1}^n \epsilon (\frac{i\bar h}{2(1+|\psi_1|^2+\psi_2^2)} \nonumber\\
& &\times(\psi_{1k} \frac{\triangle \psi_{1k}}{\epsilon}-\psi_{2k}^* \frac{\triangle \psi_{2k}}{\epsilon}-\psi_{1k}^* \frac{\triangle \psi_{1k}^*}{\epsilon}-\psi_{2k} \frac{\triangle \psi_{2k}^*}{\epsilon}) \nonumber\\
& &-\langle\psi_k|\hat H|\psi_k \rangle)) \nonumber\\
& &
\end{eqnarray}

Below is presented lagrangian L, obtained from “formal” functional integral:

\begin{eqnarray}
L=i(\frac{\bar h}{2(1+\psi_1^2+\psi_2^2)}(\psi_1^* \psi_{t1}+\psi_2^*\psi_{t2}-\psi_1\psi_{t1}^*-\psi_2\psi_{t2}^*)-\langle\psi|\hat H|\psi \rangle
\end{eqnarray}

In order to obtain classical equations of motion, the following condition should be used:

\begin{eqnarray}
0=\delta S&=& \int_t^{t{'}}(\frac{\partial L}{\partial\psi_1}\triangle \psi_1+\frac{\partial L}{\partial\psi_{t1}}\triangle \psi_{t1}+\frac{\partial L}{\partial\psi_2}\triangle \psi_2+\frac{\partial L}{\partial\psi_{t2}}\triangle \psi_{t2}+c.c.)dt \nonumber\\
&=& \int_t^{t{'}}((\frac{\partial L}{\partial\psi_1}-\frac{d}{dt}(\frac{\partial L}{\partial\psi_{t1}}))\delta \psi_1+(\frac{\partial L}{\partial\psi_2}-\frac{d}{dt}(\frac{\partial L}{\partial\psi_{t2}}))\delta \psi_2+c.c.) dt \nonumber\\
& &
\end{eqnarray}

Because variations $\delta \psi_i$  and $\delta \psi_i^*$  are independent and arbitrary, then

\begin{eqnarray}
\frac{d}{dt}(\frac{\partial L}{\partial\psi_{t1}})-\frac{\partial L}{\partial\psi_1}=0, & &{  }\frac{d}{dt}(\frac{\partial L}{\partial\psi_{t1}^*})-\frac{\partial L}{\partial\psi_1^*}=0 \nonumber\\
\frac{d}{dt}(\frac{\partial L}{\partial\psi_{t2}})-\frac{\partial L}{\partial\psi_2}=0, & &{  }\frac{d}{dt}(\frac{\partial L}{\partial\psi_{t2}^*})-\frac{\partial L}{\partial\psi_2^*}=0
\end{eqnarray}

If Lagrangian relation is used in the above equations, the classical equations take the following forms:

\begin{eqnarray}
\psi_{t1}&=&-i\frac{(1+\psi_1^2+\psi_2^2)^2}{\bar h} \frac{\partial \langle\psi|H|\psi\rangle}{\partial \psi_1^*} \nonumber\\
\psi_{t2}&=&-i\frac{(1+\psi_1^2+\psi_2^2)^2}{\bar h} \frac{\partial \langle\psi|H|\psi\rangle}{\partial \psi_2^*} \nonumber\\
\psi_{t1}^*&=&i\frac{(1+\psi_1^2+\psi_2^2)^2}{\bar h} \frac{\partial \langle\psi|H|\psi\rangle}{\partial \psi_1} \nonumber\\
\psi_{t2}^*&=&i\frac{(1+\psi_1^2+\psi_2^2)^2}{\bar h} \frac{\partial \langle\psi|H|\psi\rangle}{\partial \psi_2} 
\end{eqnarray}
These are classical equations in complex variables in SU(3) group.

\section{Properties of the SU(4) coherent state and classical dynamics}

The SU(4) CS is written in the following form

\begin{eqnarray}
|\psi\rangle=exp(\sum_{i=1}^{3}(\xi_i T_i^+-\bar \xi_i T_i^-))|0\rangle=(1+\sum_i^3|\psi_i|^2)^{-1/2}(|0\rangle+\sum_i^3\psi_i|i\rangle)
\end{eqnarray}

Where $ T_i$  are generators of SU(4) group that related to the following 15 matrices. 

\begin{eqnarray}
\beta_1 &=&
 \left(
\begin{array}{cccc}
0 & 0&0&0 \\
0 & 0&0&0 \\
0&0&0&-i  \\
0&0&i&0
\end{array}
\right), \;
\beta_2=
 \left(
\begin{array}{cccc}
0 & 0 &0&0\\
0& 0 &0&-i \\
0& 0 &0&0\\
0&i&0&0 
\end{array}
\right),\; \nonumber\\
\beta_3&=&
 \left(
\begin{array}{cccc}
0&0&0&-i \\
0&0&0&0 \\
0&0&0&0\\
i&0&0&0
\end{array}
\right), \;
\beta_4=
 \left(
\begin{array}{cccc}
0&0&0&0 \\
0&0&-i&0 \\
0&i&0&0\\
0&0&0&0
\end{array}
\right), \; \nonumber\\
\beta_5&=&
 \left(
\begin{array}{cccc}
0&0&-i&0 \\
0&0&0&0\\
i&0&0&0\\
0&0&0&0
\end{array}
\right), \;
\beta_6=
 \left(
\begin{array}{cccc}
0&-i&0&0 \\
i&0&0&0 \\
0&0&0&0\\
0&0&0&0
\end{array}
\right), \; \nonumber\\
\beta_7&=&
 \left(
\begin{array}{cccc}
0&0&0&0\\
0&0&0&0 \\
0&0&0&1\\
0&0&1&0
\end{array}
\right),\; 
\beta_8=
 \left(
\begin{array}{cccc}
0&0&0&0 \\
0&0&0&1 \\
0&0&0&0\\
0&1&0&0 
\end{array}
\right), \; \nonumber\\
\beta_9&=&
 \left(
\begin{array}{cccc}
0&0&0&1\\
0&0&0&0 \\
0&0&0&0\\
1&0&0&0
\end{array}
\right),\; 
\beta_{10}=
 \left(
\begin{array}{cccc}
0&0&0&0 \\
0&0&1&0 \\
0&1&0&0\\
0&0&0&0 
\end{array}
\right), \; \nonumber\\
\beta_{11}&=&
 \left(
\begin{array}{cccc}
0&0&1&0\\
0&0&0&0 \\
1&0&0&0\\
0&0&0&0
\end{array}
\right),\; 
\beta_{12}=
 \left(
\begin{array}{cccc}
0&1&0&0 \\
1&0&0&0 \\
0&0&0&0\\
0&0&0&0 
\end{array}
\right), \; \nonumber\\
\beta_{13}&=&
 \left(
\begin{array}{cccc}
1&0&0&0\\
0&-1&0&0 \\
0&0&0&0\\
0&0&0&0
\end{array}
\right),\; 
\beta_{14}=\frac{1}{\sqrt{3}}
 \left(
\begin{array}{cccc}
1&0&0&0 \\
0&-1&0&0 \\
0&0&-2&0\\
0&0&0&0 
\end{array}
\right), \; \nonumber\\
\beta_{15}&=&\frac{1}{\sqrt{6}}
 \left(
\begin{array}{cccc}
1&0&0&0\\
0&-1&0&0 \\
0&0&1&0\\
0&0&0&-3
\end{array}
\right),\; 
\end{eqnarray}

Coherent state is 

\begin{eqnarray}
|\psi\rangle=(1+\psi_1^2+\psi_2^2+\psi_3^2)^{1/2}(|0\rangle+\psi_1|1\rangle+\psi_2|2\rangle+\psi_3|3\rangle)
\end{eqnarray}

These states are parameterized by three complex functions  $\psi_1$ , $\psi_2$  and $\psi_3$ so the system lived on a sex-dimensional real manifold. Where 

\begin{eqnarray}
\psi_i=\frac{\xi_i}{|\xi|}tan|\xi| &  & {  } |\xi|=\sqrt{\sum_{i=1}^3 |\xi_i|^2}, i=1,2,3
\end{eqnarray}

the Casimir operator and averaged are

\begin{eqnarray}
\hat C_2&=&(S^z)^2+\frac{1}{2}(S^+S^-+S^-S^+)=Q^{zz}+\frac{1}{2}(Q^{+-}+Q^{-+}) \nonumber\\
\hat C_2&=&s(s+1)\hat I,for  s=3/2
\end{eqnarray}

 Similar to SU(2) and SU(3) groups, transition amplitude given in the following form:
\begin{eqnarray}
T(\psi^{'},t^{'};\psi,t)&=& \nonumber\\
& &lim_{n\rightarrow\infty}\int...\int\prod_{k=1}^{n-1} d\mu(\psi_k) exp(\frac{i}{\bar h}\sum_{k=1}^n \epsilon (\frac{i\bar h}{2(1+|\psi_1|^2+\psi_2^2+\psi_3^2)} \nonumber\\
& &\times(\psi_{1k} \frac{\triangle \psi_{1k}}{\epsilon}+\psi_{2k}^* \frac{\triangle \psi_{2k}}{\epsilon}+\psi_{3k}^* \frac{\triangle \psi_{3k}}{\epsilon}-\psi_{1k}^* \frac{\triangle \psi_{1k}^*}{\epsilon}-\psi_{2k} \frac{\triangle \psi_{2k}^*}{\epsilon} \nonumber\\
& &-\psi_{3k} \frac{\triangle \psi_{3k}^*}{\epsilon})-\langle\psi_k|\hat H|\psi_k \rangle)) \nonumber\\
& &
\end{eqnarray}

Below is presented lagrangian L, obtained from “formal” functional integral:

\begin{eqnarray}
L&=&i(\frac{\bar h}{2(1+\psi_1^2+\psi_2^2+\psi_3^2)})(\psi_1^* \psi_{t1}+\psi_2^*\psi_{t2}+\psi_3^*\psi_{t3}-\psi_1\psi_{t1}^*-\psi_2\psi_{t2}^*-\psi_3\psi_{t3}^*) \nonumber\\
& &-\langle\psi|\hat H|\psi \rangle
\end{eqnarray}

In order to obtain classical equations of motion, the following condition should be used:

\begin{eqnarray}
0=\delta S&=& \int_t^{t{'}}(\frac{\partial L}{\partial\psi_1}\triangle \psi_1+\frac{\partial L}{\partial\psi_{t1}}\triangle \psi_{t1}+\frac{\partial L}{\partial\psi_2}\triangle \psi_2 +\frac{\partial L}{\partial\psi_{t2}}\triangle \psi_{t2} \nonumber\\
& &+\frac{\partial L}{\partial\psi_3}\triangle \psi_3+\frac{\partial L}{\partial\psi_{t3}}\triangle \psi_{t3}+c.c.)dt  \nonumber\\
&=& \int_t^{t{'}}((\frac{\partial L}{\partial\psi_1}-\frac{d}{dt}(\frac{\partial L}{\partial\psi_{t1}}))\delta \psi_1+(\frac{\partial L}{\partial\psi_2}-\frac{d}{dt}(\frac{\partial L}{\partial\psi_{t2}}))\delta \psi_2 \nonumber\\
& &+(\frac{\partial L}{\partial\psi_3}-\frac{d}{dt}(\frac{\partial L}{\partial\psi_{t3}}))\delta \psi_3+c.c.) dt \nonumber\\
& &
\end{eqnarray}

Because variations $\delta \psi_i$  and $\delta \psi_i^*$  are independent and arbitrary, then

\begin{eqnarray}
\frac{d}{dt}(\frac{\partial L}{\partial\psi_{t1}})-\frac{\partial L}{\partial\psi_1}=0, & &{  }\frac{d}{dt}(\frac{\partial L}{\partial\psi_{t1}^*})-\frac{\partial L}{\partial\psi_1^*}=0 \nonumber\\
\frac{d}{dt}(\frac{\partial L}{\partial\psi_{t2}})-\frac{\partial L}{\partial\psi_2}=0, & &{  }\frac{d}{dt}(\frac{\partial L}{\partial\psi_{t2}^*})-\frac{\partial L}{\partial\psi_2^*}=0 \nonumber\\
\frac{d}{dt}(\frac{\partial L}{\partial\psi_{t3}})-\frac{\partial L}{\partial\psi_3}=0, & &{  }\frac{d}{dt}(\frac{\partial L}{\partial\psi_{t3}^*})-\frac{\partial L}{\partial\psi_3^*}=0 
\end{eqnarray}

If Lagrangian relation is used in the above equations, the classical equations take the following forms:

\begin{eqnarray}
\psi_{t1}&=&-i\frac{(1+\psi_1^2+\psi_2^2+\psi_3^2)^2}{\bar h} \frac{\partial \langle\psi|H|\psi\rangle}{\partial \psi_1^*} \nonumber\\
\psi_{t2}&=&-i\frac{(1+\psi_1^2+\psi_2^2+\psi_3^2)^2}{\bar h} \frac{\partial \langle\psi|H|\psi\rangle}{\partial \psi_2^*} \nonumber\\
\psi_{t3}&=&-i\frac{(1+\psi_1^2+\psi_2^2+\psi_3^2)^2}{\bar h} \frac{\partial \langle\psi|H|\psi\rangle}{\partial \psi_3^*} \nonumber\\
\psi_{t1}^*&=&i\frac{(1+\psi_1^2+\psi_2^2+\psi_3^2)^2}{\bar h} \frac{\partial \langle\psi|H|\psi\rangle}{\partial \psi_1} \nonumber\\
\psi_{t2}^*&=&i\frac{(1+\psi_1^2+\psi_2^2+\psi_3^2)^2}{\bar h} \frac{\partial \langle\psi|H|\psi\rangle}{\partial \psi_2} \nonumber\\
\psi_{t3}^*&=&i\frac{(1+\psi_1^2+\psi_2^2+\psi_3^2)^2}{\bar h} \frac{\partial \langle\psi|H|\psi\rangle}{\partial \psi_3}
\end{eqnarray}
These equations are classical equations in complex variables in SU(4) group.

\section{Properties of the SU(2S+1) coherent state and classical dynamics}

The SU(2S+1) CS for $S\geq1$  is  written in the following form:

\begin{eqnarray}
|\psi\rangle=exp(\sum_{i=1}^{2S}(\xi_i T_i^+-\bar \xi_i T_i^-))|0\rangle=(1+\sum_i^{2S}|\psi_i|^2)^{-1/2}(|0\rangle+\sum_i^{2S}\psi_i|i\rangle)
\end{eqnarray}

Where $T_i$ are generators of SU(2S+1) or SU(n) group. These generators can be represented by  $(n^2-n)$  off-diagonal matrices and $(n-1)$  diagonal matrices. We take $ e_j^h$   as a basis for the group SU(n) and Non-diagonal element of this basis are[4]

\begin{eqnarray}
\beta_j^h=-i(e_j^h-e_h^j),& & {  }\Theta_j^h=e_j^h+e_h^j, 1\le h< j\le n
\end{eqnarray}

The diagonal elements are:

\begin{eqnarray}
\eta_m^n=\sqrt{\frac{2}{m(m+1)}}(\sum_{j=1}^m e_j^j-me_{m+1}^{m+1}) & & {  }1\le m\le n-1
\end{eqnarray}

These states are parameterized by complex functions $\psi_i$ . Where 

\begin{eqnarray}
\psi_i=\frac{\xi_i}{|\xi|}tan|\xi| &  & {  } |\xi|=\sqrt{\sum_{i=1}^{2S} |\xi_i|^2}
\end{eqnarray}

the Casimir operator and averaged are:

\begin{eqnarray}
\hat C_2&=&(S^z)^2+\frac{1}{2}(S^+S^-+S^-S^+)=Q^{zz}+\frac{1}{2}(Q^{+-}+Q^{-+}) \nonumber\\
\hat C_2&=&s(s+1)\hat I
\end{eqnarray}

 The transition amplitude (propagator) from state $ |\psi \rangle$  at time t to the state $|\psi^{'}\rangle$   at time $ t^{'}$   is given by 

\begin{eqnarray}
T(\psi^{'},t^{'},\psi,t)&=&\langle\psi^{'}|exp(-\frac{i}{\bar h}\hat H (t^{'}-t))|\psi \rangle \nonumber\\
&= &lim_{n\rightarrow\infty}\int...\int\prod_{k=1}^{n-1} d\mu(\psi_k) exp(\frac{i}{\bar h}\sum_{k=1}^n \epsilon (\frac{i\bar h}{2(1+\sum_i^{2S}\psi_i^2)} \nonumber\\
& &\times(\sum_i^{2S}\psi_{ik} \frac{\triangle \psi_{ik}^*}{\epsilon}-\sum_i^{2S}\psi_{ik}^* \frac{\triangle \psi_{ik}}{\epsilon})-\langle\psi_k|\hat H|\psi_k \rangle) \nonumber\\
& &
\end{eqnarray}

Below is presented the $“formal” $ functional integral of the above expression:

\begin{eqnarray}
T&=&\int d\mu(\psi)exp(\frac{i}{\bar h}S) \nonumber\\
S&=& \int_t^{t{'}}L(\psi_i(t), \psi_{ti} (t), \psi_i^*(t), \psi_{ti}^*(t))dt , i=1..2S
\end{eqnarray}

Then lagrangian L is given by 

\begin{eqnarray}
L&=&i(\frac{\bar h}{2(1+\sum_i^{2S}\psi_i^2)})(\sum_i^{2S}\psi_i^* \psi_{ti}-\sum_i^{2S}\psi_i \psi_{ti}^*)-\langle\psi|\hat H|\psi \rangle
\end{eqnarray}

In order to obtain classical equation, the following condition should be used:

\begin{eqnarray}
0=\delta S&=& \int_t^{t{'}}( \sum_i^{2S}\frac{\partial L}{\partial\psi_i}\triangle \psi_i+\sum_i^{2S}\frac{\partial L}{\partial\psi_{ti}}\triangle \psi_{ti}+c.c.)dt  \nonumber\\
&=& \int_t^{t{'}}(\sum_i^{2S}(\frac{\partial L}{\partial\psi_i}-\frac{d}{dt}(\frac{\partial L}{\partial\psi_{ti}}))\delta \psi_i+c.c.) dt \nonumber\\
& &
\end{eqnarray}

Because variations $\delta \psi_i$  and $\delta \psi_i^*$  are independent and arbitrary, then

\begin{eqnarray}
\frac{d}{dt}(\frac{\partial L}{\partial\psi_{ti}})-\frac{\partial L}{\partial\psi_i}=0, & &{  }\frac{d}{dt}(\frac{\partial L}{\partial\psi_{ti}^*})-\frac{\partial L}{\partial\psi_i^*}=0 \nonumber\\
\end{eqnarray}

If L relation is used in the above equations, the classical equations take the following forms:

\begin{eqnarray}
\psi_{ti}&=&-i\frac{(1+\sum_i^{2S}\psi_i^2)}{\bar h}\frac{\partial \langle \psi |H|\psi \rangle}{\partial \psi_i^*} \nonumber\\
\psi_{ti}&=&i\frac{(1+\sum_i^{2S}\psi_i^2)}{\bar h}\frac{\partial \langle \psi |H|\psi \rangle}{\partial \psi_i}, i=1..2S
\end{eqnarray}

\section{Discussion}

Our formulation can be used to write  the field theory for the SU(n) Heisenberg or Non-Heisenberg   model and to study its spectrum and topological aspects. If Hamiltonian is used in above equations, nonlinear equations describing properties of systems are obtained.

For example, this formalizatin can be used in the dynamical description of magnetic substances with spins $S\geq1/2$ . It is shown that the minimum number of dynamical variables (and, consequently, of equations for them) necessary to consider all the interactions allowed by the magnitude of the spin adequately is equal to 4S. A set of 4S equations that describe the dynamics of an magnetic material system are explicitly derived on the basis of the single- site coherent states for the SU(2S+1) Lie group. Those physical situations are considered whose most important feature is not the orientational motion of the magnetization vector, but the dynamics of the multipole degrees of freedom, which constitute an important element of the total dynamics.

In analyzing SU(2) group, there is only dipole moment, and the length of magnetization vector is constant but in SU(3) group there are both dipole and quadrupole moment [5]. Generally, in SU(n) group, there are  dipole up to multipole moments that must be considered.

There are 3 generators in SU(2) group (Pauli matrices),  8 generators in SU(3) group, 3 of which related to dipole moment and  the rest are related to quadrupole moment. In general , in SU(n) group there are $n^2-1$   generators that form multipole moments.


\begin{thebibliography}{9}
\bibitem{law}	A. Perelomov. Generalized coherent states and applications. Nauka, Mascow, 1987.
\bibitem{law}	Path integral in the representation of SU(2) coherent state and classical dynamics in a generalized phase space, H. Kuratsuji and T. Suzuki, J.Math. Phys. 21(3) 1980.
\bibitem{law}	Generalized coherent states and spin system, V.G.Makhankov, arxiv: chao-dyn/9602009V1 8 feb 1996.
\bibitem{law}	Generalized coherent states for SU(n) systems, Kae Nemoto, arXiv:quant-ph/0004087v1, 22 apr 2000.
\bibitem{law}	N. A. Mikushina and A. S. Moskivn, Phys. Lett A, V. 302, Issue 1, 8-16 , 2002.
\end{thebibliography}
\end{document}